\documentstyle[twocolumn,prc,aps,epsfig,psfig]{revtex}

     \newcommand{\pathnow}{}
\begin{document}\hbadness=10000\topmargin -1cm
  \hyphenation{strang-en-ess}
\twocolumn[\hsize\textwidth\columnwidth\hsize\csname %
@twocolumnfalse\endcsname
\title{Rapidity particle spectra in sudden hadronization of QGP
}
\author{Jean Letessier and Johann Rafelski}
\address{
 Laboratoire de Physique Th\'eorique et Hautes  Energies,\\
Universit\'e Paris 7, 2 place Jussieu, F--75251 Cedex 05\\[0.2cm]
and\\[0.2cm]
Department of Physics, University of Arizona, Tucson, AZ 85721
}
\date{June 2001}
\maketitle
\begin{abstract}
We show that  the remaining internal longitudinal flow of colliding 
quarks in nuclei offers a natural explanation for the diversity of 
rapidity spectral shapes observed in Pb--Pb 158$A$ GeV 
nuclear collisions. Thus QGP sudden hadronization reaction picture
is a suitable approach to explain the  rapidity spectra of hadrons 
produced.  \\

\end{abstract}
]
\begin{narrowtext}
If a quark-gluon plasma deconfined state super cools and
decomposes in a sudden fashion into free streaming hadrons
\cite{Raf00}, we should be able to see 
in the hadron spectra certain features of quark statistical
distributions. Such a QGP fingerprint on $m_\bot$-spectra
of strange baryons and antibaryons which are seen to be
identical is one of key 
indicators for the sudden hadronization picture \cite{TR01}.

A question which has remained open  is 
how sudden hadronization model can be reconciled with the
diverse shapes of 
rapidity spectra of observed hadrons. Great differences are known to
be present: while hadrons made of quarks brought into
the collision, such as, e.g., protons, show a pronounced double 
hump structure, progressively more peaked distributions are
seen for particles which are more and more composed of
newly made quarks \cite{App99,App98,Afa00}. 

We present a schematic description of this phenomenon
based on the assumption that the quark momentum distribution
in QGP is imprinted on hadrons produced, and not erased in a
post hadronization period. The model we explore is a direct 
continuation of the study of $m_\bot$ shape and abundance of 
hadrons produced \cite{Let00}.

We note that many  light quarks are brought into the collision 
fireball by  colliding nuclei --- 
in  158$A$ GeV  Pb--Pb central interactions there
are 1000\,+ such valance quark participants and
there is a similar total number of hadrons produced. 
In general the two incoming quark fluids can retain
memory  of the early stage in the collision: in the center of momentum
frame two quark fluids collide with initial rapidities $y_i=\pm2.92$.
In order to reproduce the proton data,  we 
assume in what follows,  that
the {\it average} remaining longitudinal quark flow is $y_f=\pm1$. 
This average includes quarks from $q\bar q$ pairs 
produced, which are mostly found at central rapidity. 
We further assume that 
the average flow of $\bar s, s,\bar q$ quarks
is zero. 

If quarks were not confined, their emission spectrum
would be computed in a thermal emission model with
flow.  We proceed to establish this spectrum. 
Each emitting volume element is assumed to be
in motion, with a four-vector $u^\mu$, and in the local frame
at rest the temperature in the volume element is $T$, and the 
local quark distribution is of Boltzmann type. 
In order to count the  states, we apply Touschek invariant 
phase space measure for the level density \cite{Tou68,Hag78}:
\begin{equation}\label{tou}
\frac{V_0d^3p}{2\pi^3} e^{-E/T}\to \frac{V_\mu p^\mu}{(2\pi)^3}
d^4p\,2\delta_0(p^2-m^2)e^{-p_\mu u^\mu/T}\,.
\end{equation}
Here, $ V_\mu=V_0u_\mu$ is the flowing volume element as observed in the
laboratory frame. $V_0$ is the comoving volume element in the local
rest frame.  $\delta_0$ is the Dirac-delta function for the positive
roots only. This is an invariant function for all proper Lorentz transformations. 
Using the rapidity and 
transverse momentum  $p_\bot$ (or mass $m_\bot=\sqrt{m^2+p^2_\bot}$) 
variables of a particle in a cylindrical symmetric  decomposition with respect 
to the collision axis
\begin{equation}\label{pcyl}
p^\mu=(m_\bot \cosh y, p_\bot \cos \phi_p,  p_\bot \sin \phi_p, m_\bot \sinh y)\,,
\end{equation}
 we recover the well known invariant measure in Eq.\,(\ref{tou}):
\begin{eqnarray}\nonumber
2\delta_0(p^2-m^2)d^4p=\frac{d^3p}{E}&=&m_\bot dm_\bot dy d\phi_p\\
&=&p_\bot dp_\bot dy d\phi_p\,.\label{invmeas}
\end{eqnarray}\label{u4cyl}
The four-velocity $u^\mu$ allowing for both  transverse and longitudinal flows is:
\begin{eqnarray}\label{umucyl}
u^\mu&=&(u_0,u_x,u_y,u_z)\,,\quad u^2\,=1\,;\\
u_0&=&\cosh y_\bot \cosh y_\parallel\,;\quad
u_x= \sinh y_\bot \cos \phi_u\,;\nonumber \\
u_y&=& \sinh y_\bot\sin \phi_u\,;\quad\,\
u_z= \cosh y_\bot \sinh y_\parallel\,.\nonumber
\end{eqnarray}
We recall the usual relations:
$$
\frac{m_\bot}m=\gamma_\bot =\cosh y_\bot\,, \quad
\frac{p_\bot}m=v_\bot \gamma_\bot =\sinh y_\bot\,, 
$$ 
which illustrate the structural identity of Eq.\,(\ref{pcyl}) with  Eq.\,(\ref{umucyl}).

It is straightforward now to obtain the quantity required to construct spectra 
 Eq.\,(\ref{tou})\,,
\begin{equation}\label{upcyl}
u_\mu p^\mu=\gamma_\bot\left[
 m_\bot\cosh(y-y_\parallel)-p_\bot v_\bot \cos\phi \right]\,,
\end{equation}
where $\phi=\phi_p-\phi_u$. For a suitable choice of coordinate 
system in which the $x$-axis is pointing in the direction of 
transverse  flow vector, $\phi_u=0$, 
and we can use the particle emission angle as the azimuthal angle of 
integration for radiation emitted from the volume element: $\phi=\phi_p$. This has
the advantage that we can easily restrict the counting of particles
to those emitted {\it outwards} including $0<\phi_p\le \pi/2$
instead of the full range $0<\phi_p\le \pi$

In Eq.\,(\ref{upcyl}), the emitted particle variables are as usual $y,m_\bot$
and the flow vector variables used are $y_\parallel$ and $v_\bot$ with
$\gamma_\bot=1/\sqrt{1-v_\bot^2}$. 
The invariant spectrum written explicitly is:
\begin{eqnarray}
\frac{d^2N}{m_\bot dm_\bot dy}&=&\!\int\!\frac{d\phi \,\gamma_\bot}{(2\pi)^3}
\left[
 m_\bot\cosh(y-y_\parallel)-p_\bot v_\bot \cos\phi \right]\nonumber\\[0.3cm]
&&\hspace*{-0.3cm}\times\ e^{-\gamma_\bot\left[
 m_\bot\cosh(y-y_\parallel)-p_\bot v_\bot \cos\phi \right]/T}\,.\label{specinv}
\end{eqnarray}
There are some differences in detail between this traditional 
expression \cite{Tou68,Hag78}
and recent proposals based on consideration of the
freeze-out surface geometric properties \cite{Sch95}. Certain features 
we find are the same in both approaches: the rapidity $y$
distribution is not appreciably influenced by transverse dynamics (i.e., 
transverse velocity and temperature), which effectively separates.  
The $m_\bot$ spectra are only dependent on $T,v_\bot$, but practically 
not on the longitudinal flow $y_\parallel$. We note that
Eq.\,(\ref{specinv}) retains absolute  number of particles 
independent of the flow parameters, if integral over $\phi$
includes the full range $0\le\phi\le \pi$. 

The rapidity spectra of massless quanta in the deconfined phase are 
shown in figure \ref{rapidcomp}. We see how the thermal spectrum 
($m=0$, $T=145$ MeV, $v_\bot=0.52$) gradually `flows' apart as $y_\parallel$
is increased from $y_\parallel=0$ (dotted) to $y_\parallel=1.5$ (solid) 
in step of 0.5\,. Included in the double numerical integral 
(over $\phi$ and $m_\bot$) are particles emitted outwards with reference to 
the direction of the flow vector of the  volume element. 
Model studies of kinetic dynamics show that 
 inside the QGP matter we find 
a mix of fluids, the incoming quarks are still retaining some of the original flow 
while all newly made quark pairs have no memory  of the initial condition
of colliding matter. In particular, strange quark pairs made in the plasma do not
flow. 

Final state hadrons are born from deconfined quarks with such
spectra.  In the previously studied case of $m_\bot$ spectra \cite{TR01} 
it was found that the transverse flow parameters of the hadronizing quark 
matter were in agreement with  results of 
chemical particle production parameters. Thus the shape of $m_\bot$ spectra
of hyperons, antihyperons and kaons was as if these particles
were directly made of hadronizing QGP quanta. 
Addressing here  the rapidity spectra we will take this point 
of view but proceed more carefully.
As hyperon are formed at the fireball breakup, any remaining longitudinal
flow present among fireball constituents 
will be imposed on the product particle, thus $\Lambda$-spectra
 containing potentially two original valence quarks are stretched in
$y$, which  $\overline\Lambda$-$y$-spectra are not, as they are made from newly 
formed particles. One would thus expect that anti-hyperons comprising 
only newly made quarks will appear without flow, while $\Lambda$-spectra
will show flow, if such was still present.

\begin{figure}[tb]
\vspace*{-3.2cm}
\centerline{\hspace*{.8cm}
\psfig{width=9.5cm,clip=,figure=\pathnow 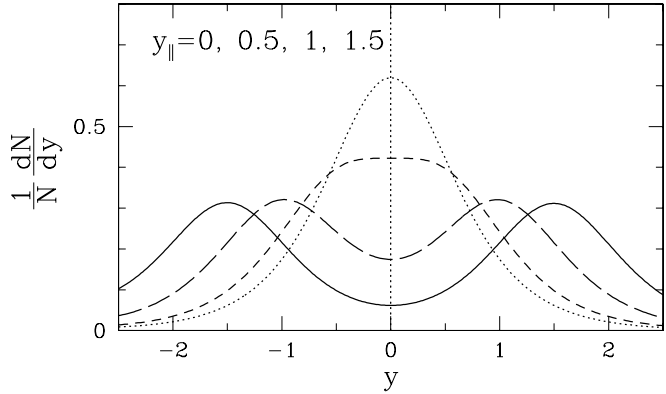}
}
\vspace*{-0.4cm}
\caption{ 
Rapidity spectra with flow of massless QGP quanta:
dotted: no flow $y_\parallel=0$, dashed, $y_\parallel=0.5$, longdashed 
$y_\parallel=1$, solid, $y_\parallel=1.5$.
\label{rapidcomp}
}
\end{figure}

We thus implement a particle-staged turn-on of longitudinal flow at SPS where 
in central Pb--Pb collisions the incoming valance quark number and 
total number of hadrons produced is about the same. For
protons we  assume that effectively all quarks are remembering some of the
original flow. For  $\Lambda$ with one strange quark we take 
a mix of 2/3 weight in the  spectrum with flow and 1/3 without. 
For particles like $K^+(u\bar s)$  we take a 50-50\% mix and for
all newly made particles like $\overline\Lambda, \Phi$ we 
assume that only no flow components contribute.  

To describe proton rapidity spectra we assume after inspection of 
\cite{App99} that the longitudinal flow is $y_\parallel\simeq \pm 1$,
and choose this value without a further attempt to fit the spectra.
With $m_q=0$, $T=145$ MeV and $v_\bot=0.52c$ (both parameters could be chosen
very differently without altering the result, 
these were taken in view of results of  \cite{TR01}),
we obtain the  proton rapidity spectrum  we present 
in figure \ref{specflow}, bottom panel.
We note that we have to average over positive 
and negative flows $y_\parallel=\pm1$, since
the collision in the CM frame involves both. 

The strange quark content of the $\Lambda$ contributes with 
relative strength 33\% in the central 
region, which results in a flat distribution --- see the second panel
from the bottom in figure \ref{specflow}. Next we
take a 50-50\% mix which gives
us the result shown in the third panel from the bottom, corresponding to 
the $K^+(u\bar s)$ rapidity distribution and when resonance decays are
considered, also  pion distribution. Finally, in the top panel we show prototype of 
the rapidity distribution arising for all completely newly made particles
such as $\overline\Lambda, \Phi$.  Though far from perfect, we see in this
schematic study practically exactly the behavior observed.

The same schematic rapidity behavior 
is seen in S--S collisions at 200$A$ GeV for the particles here considered.
A direct comparison of the charged particle distributions between 
Pb--Pb and S--S collision systems seen in Ref.\cite{Wie96}
suggests  that $y_\parallel$ is about 0.4 units of rapidity larger in the
lighter collision system.  Though seemingly a small change, this opens
by 50\% the gap between the fluids as we saw in figure \ref{rapidcomp}
and a much more pronounced central 
rapidity reduction in certain particle  abundances ensues.

\begin{figure}[tb]
\centerline{\hspace*{1.6cm}
\psfig{width=11.9cm,clip=,figure=\pathnow 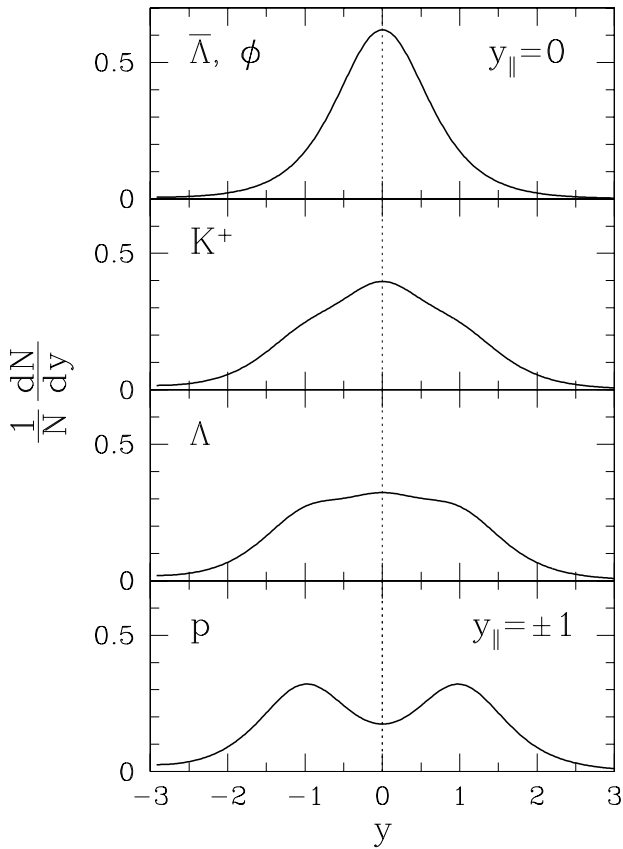}
}
\caption{ 
Rapidity particle spectra: schematic representation within a thermal 
model with flow, parameters chosen for $\sqrt{s_{NN}}=17.2$ GeV. 
See text for details.
\label{specflow}
}
\end{figure}

Our schematic  method can help understand the 
particle rapidity distributions at RHIC. The PHOBOS collaboration
has presented  a nearly flat  distribution of charged hadrons 
with small peaks near to pseudo rapidity $\eta=1.5$ at the edge
 \cite{Rol01,Bac01}. Trying our model 
we finds that rather than to have a 3-fluid quark source,
we need now a  continuous distribution of sources residing
in the range of rapidity $-1.6\le y\le 1.6$. In other words 
if the hadron rapidity distribution is as before offering
an image of the quark flow, we have a flow plateau as expected 
in the hydrodynamical scaling solution  
proposed by Bj\o rken \cite{Bjo83}.

In summary of this brief contribution, we have shown, within a schematic  
dynamic model, that quark flow in QGP can explain the diversity of particle
rapidity  spectra observed at  $\sqrt{s_{NN}}=17.2$ GeV, where clearly
we do not see yet a high energy scaling behavior.  
What we have shown  is that a three component fluid, two made of 
quarks still flowing against each other, and a third one made of 
newly made components (antiquarks, strange quarks), 
is capable to imprint on the rapidity
spectra just the right systematic behavior. This transfer of
quark dynamics to hadron dynamics is similar to what
was found  in the study of the $m_\bot$ spectra.
For hadrons to emerge with momentum spectra of quarks we need 
a hadronization with sudden breakup of the 
(super cooled) QGP plasma phase into free-streaming hadrons.
Thus this study shows that such a picture is capable to explain rapidity
spectra.

\vskip 0.3cm
\noindent{\bf Acknowledgments:}
Work supported in part by a grant from the U.S. Department of
Energy,  DE-FG03-95ER40937. Laboratoire de Physique Th\'eorique 
et Hautes Energies, University Paris 6 and 7, is supported 
by CNRS as Unit\'e Mixte de Recherche, UMR7589.


\end{narrowtext}
\end{document}